\documentclass[10 pt]{article}
\usepackage{amsmath,amsfonts,amssymb,bm}
\usepackage{float}
\usepackage{subfigure}
\usepackage{graphicx}
\usepackage{wrapfig}
\usepackage[margin=1in]{geometry}
\usepackage{hyperref}

\begin{document}
\title{On the Conventionality of Simultaneity and the Huygens-Fresnel-Miller Model of Wave Propagation}
\author{Robert D. Bock\footnote{robert at r-dex.com} \\ R-DEX Systems \\ \url{www.r-dex.com}}
\date{\today}

\maketitle

\begin{abstract}
\noindent 
We identify a fundamental space-time invariance principle by combining the thesis of the conventionality of simultaneity with the Hugyens-Fresnel-Miller model of wave propagation. By following the standard gauge prescription, we show that the classical electromagnetic potentials influence the one-way speed of light.

\end{abstract}

\noindent \textbf{KEY WORDS}:  conventionality of simultaneity, clock synchronization, Huygens-Fresnel, wave propagation of light, gauge theory, electromagnetism, relativity 
                             
\bigskip

There is a long-standing debate in the literature regarding the conventionality of simultaneity \cite{janis2014}.  On the one hand, supporters of the conventionality thesis (e.g., \cite{reichenbach1958, grunbaum1973, zhang1997,anderson1998}) advocate that clock simultaneity is an arbitrary convention that permits different one-way speeds of light.  According to this thesis, all simultaneity conventions that preserve the experimentally measured two-way speed of light are equivalent.  On the other hand, opponents of the conventionality thesis (e.g., \cite{salmon1977, malament1977, ohanian2004}) argue that standard synchrony defined by Einstein synchronization is the only clock synchronization convention that is permitted by fundamental physical laws.  Furthermore, they argue that the one-way speed of light can be measured independently of the synchronization convention and is equal to the experimentally measured two-way speed of light.  Although this topic has received significant attention throughout the years this debate remains unsettled.  The absence of indisputable experimental evidence in favor of either interpretation has contributed to the prolongation of the debate.

To the author's knowledge, arguments both supporting and refuting the conventionality of simultaneity ignore the wavelike nature of light propagation. Although Einstein \cite{einstein1905} refers to the spherical propagation of a wavefront for defining simultaneity in inertial frames, it is the geometrical (linear) propagation of light that is tacitly assumed in his numerous thought experiments. As is well known, Einstein chose a synchrony convention that yields a one-way speed of light that is identical to the measured two-way speed of light, recognizing that this choice is a matter of convention. More recent treatments of synchrony transformations (e.g., \cite{anderson1998}) assume that the one-way speed of light in any direction can be changed by the coordinate transformation:
\begin{equation}
\label{eq:synchronization_transformations}
x^{\prime\mu} = x^{\mu} + \delta_0^\mu b_M x^{M},
\end{equation}
where\footnote{Greek indices and lowercase Latin indices run from ($0\ldots3$).  Uppercase Latin indices run from ($1\ldots3$).} $b_M$ represent 3 real parameters. 

However, the wavelike nature of light is physically understood in the context of the Huygens-Frensel principal, such that the propagation of each point on a wave front represents a source of spherical waves (or spatiotemporal dipoles according to Miller \cite{miller1991}). According to Huygens-Fresnel and Miller, the fundamental unit for defining wave propagation is a spherical point source. In the case of the Huygens-Fresnel interpretation, the backward propagation must be neglected arbitrarily; in the case of Miller's interpretation, a second spherical point source is appropriately positioned and delayed to cancel out the backward propagation. In either case, the transformation of the one way speed of light according to the wave theory of light is not achieved by transformation (\ref{eq:synchronization_transformations}) but is achieved by transforming the local time at every point $P$ on a wave front according to:
\begin{equation}
\label{eq:synchronization_transformations_2}
t_P^{\prime} = t_P -  b\frac{r}{c},
\end{equation} 
where $r$ is the (local) radial distance from an arbitrary point on the wavefront and the secondary wavefronts and $b$ is an arbitrary constant. The local (radial) speed of light at each spherical source is given by:
\begin{equation}
c_{\pm}= \frac{c}{1\mp b},
\end{equation}
where $+$ ($-$) refers to outward (inward) radial propagation in the local spherical coordinates. The change in the speed of light induced by transformation (\ref{eq:synchronization_transformations_2}) on an Einstein-synchronized space-time is given by
\begin{equation}
\Delta c_{\pm} = \left( \frac{\mp b}{1 \mp b} \right) c.
\end{equation}
For $b\ll 1$ the change in the speed of light due to transformation (\ref{eq:synchronization_transformations_2}) becomes:
\begin{equation}
\Delta c_{\pm} = {\mp} bc.
\end{equation}
Whereas transformation (\ref{eq:synchronization_transformations}) introduces spatial anisotropy into the propagation of light, transformation (\ref{eq:synchronization_transformations_2}) does not.

According to the conventionality of simultaneity, the one-way velocity of light cannot be measured. Therefore, by combining  thesis of the conventionality of simultaneity with the Hugyens-Fresnel-Miller model of wave propagation, we are led to the following fundamental invariance principle: The laws of physics are invariant under transformation (\ref{eq:synchronization_transformations_2}) with $b$ constant, namely, the laws of physics are invariant under spherically symmetric transformations of the one-way speed of light at each point in space. Note that $r$ is not a global variable, but is defined only locally in the local coordinate systems that define the propagation from each spherical point source.

As is well known, the Einstein-Cartan-Kibble-Sciama (ECKS) theory is a gauge theory of the Poincar\'e group \cite{kibble1961} (see also \cite{utiyama1956}). The traditional theory of general relativity is a subset of ECKS gravity that follows naturally by gauging only the translation group. The additional spin coefficients emerge in ECKS because of the invariance of physical laws under local Lorentz rotations of the vierbein fields. 

Similarly, the thesis of the conventionality of simultaneity in the context of the Huygens-Fresnel-Miller model of wave propagation forces us to introduce a new gauge field, $B_\mu$, that preserves the local synchrony transformations when $b$ becomes an arbitrary function of the space-time variables. We consider a Lagrangian that is a function of a set of field variables, $\chi (x^\mu)$, and the coordinates $x^\mu$:
\begin{equation}
L\equiv L\left\{ \chi (x^\mu), \chi_{,\mu}, x^{\mu}   \right\},
\end{equation}
where $\chi_{,\mu} \equiv \partial_\mu \chi$. The variations of the coordinates and field variables under an infinitesimal transformation are:
\begin{eqnarray}
x^{\mu}& \rightarrow & x^{\prime\mu} = x^{\mu}+\delta x^{\mu} \nonumber \\
\chi(x^{\mu}) & \rightarrow &\chi^{\prime}(x^{\prime\mu})=\chi(x^{\mu})+\delta\chi(x^{\mu}).
\end{eqnarray}
We consider infinitesimal synchrony transformations:
\begin{equation}
\label{eq:synchronization_transformations_3}
\delta x^{\prime\mu} = \delta_0^\mu b r \;\;\; \delta\chi=bW\chi,
\end{equation}
where $b$ represents a real infinitesimal parameter and $W$ is the generator of the spherically-symmetric synchrony group.

According to the gauge prescription one assumes the action is invariant under a transformation group for constant parameters and then covariant derivatives are introduced to retain invariance when the parameters of the group become arbitrary functions of the coordinates.  To preserve invariance of the action under generalized synchrony transformations, we must replace the derivative $\chi_{,\mu}$ with a covariant derivative, $\chi_{;\mu}$, according to:
\begin{equation}
\chi_{;\mu}\equiv \chi_{,\mu}+B_{\mu}W\chi,
\end{equation}
This new field transforms under synchrony transformations as:
\begin{equation}
\delta B_{\mu} = -b_{,\mu}
\end{equation}
to retain invariance of the action when the parameters of the synchrony group become arbitrary functions of the coordinates. We write the Lagrangian for the free fields as 
\begin{equation}
L_0 =  - \frac{1}{4}F_0,
\end{equation}
where $F_0 = F_{\mu\nu}F^{\mu\nu}$ and ${F}_{\mu\nu}  = B_{\mu,\nu} - B_{\nu,\mu}$ is calculated from the commutator of $b$-covariant derivatives.  This produces the following field equations:
\begin{equation}
\label{eq:synchrony_field_equations}
 F_{\;\;\; ;\nu}^{\mu\nu}   = J^{\mu},
\end{equation}
where $J^{\mu} \equiv -\partial L/\partial B_{\mu}$ and $L$ is the Lagrangian that is a function of a set of fields.  We identify $B_\mu$ as being proportional to the electromagnetic potentials, $\phi_\mu$:
\begin{equation}
\label{eq:potential_def}
B_\mu = \alpha\phi_\mu,
\end{equation}
where $\alpha$ is a constant.  

Whereas Kibble demonstrated the fundamental relationship between the gravitational field and the invariance properties of the Lagrangian under the 10-parameter Poincar\'e group, we see that the electromagnetic field follows from the 5-parameter group that includes space-time translations and local, spherically-symmetric synchrony transformations. In the same way that Einstein's equivalence principle serves as the physical guide that yields torsion in the Einstein-Cartan-Kibble theory, the Huygen's-Fresnel-Miller principal along with the conventionality of simultaneity serves as the physical guide that produces the electromagnetic field in a space-time theory that is invariant under local translations. Note that this prescription successfully combines the gravitational field with the electromagnetic field, but does not incorporate the spin coefficients since Lorentz transformations do not form a group with synchrony transformations.

We conclude that the electromagnetic field modifies the one-way speed of light. Whereas the one-way speed of light remains unobservable, the relative change in the one-way speed of light that results from the presence of non-zero electromagnetic potentials can be observed. For example, the presence of non-zero electromagnetic potentials can account for many of the observed phenomena that variable speed of light (VSL) theories have attempted to explain. Indeed, a non-zero electric potential in the early universe can account for a wide range of observed phenomena consistent with VSL theories.

Let us consider a simple experiment in which light propagates along a length $L$ from an initial point $P$ to a final point $Q$. We are free to choose a synchrony convention (Einstein's convention) such that the one-way speed of light from $P$ to $Q$ is equal to the two-way speed of light. Once this synchronization convention is established we introduce an electromagnetic field, $\phi_\mu$. The change in the speed of light due to the introduction of the electromagnetic field is given by:
\begin{equation}
\Delta c= \frac{ \alpha \phi_\mu\Delta x^\mu}{1- \alpha \phi_\mu\Delta x^\mu} c.
\end{equation}
For example, consider the introduction of a constant electric potential, $\phi_0$, in a region of space for which Einstein synchronization has been established. Whereas the one-way speed of light is unobservable and defined only by the adopted convention, the relative change in the one-way speed of light when a non-zero constant electric potential $\phi_0$ is introduced can be measured:
\begin{equation}
\Delta c =\frac{ \alpha \phi_0\Delta t}{1-\alpha \phi_0\Delta t}c.
\end{equation}
For a weak field, $\Delta c = \alpha \phi_0\Delta t c$. The measurement of the relative change in the speed of light due to the introduction of an electromagnetic field in a region of space for which a synchronization convention has been established can serve to verify the proposed relationship between the electromagnetic potentials and local spherically symmetric synchrony transformations as well as to determine the value of the proportionality constant $\alpha$ in Equation (\ref{eq:potential_def}).

Let us also consider the case where two separate light beams propagate in vacuum and combine after traveling equal distances. We further assume that along the first path a constant and weak electric potential $\phi_0$ is introduced. Whereas the associated electric field vanishes, $\phi_0$ will manifest itself by producing an evolving interference pattern due to the changing relative speed of the wavefront along the path subject to $\phi_0$, according to:
\begin{equation}
\frac{\Delta c}{\Delta t} = \alpha c\phi_0.
\end{equation}

\bibliographystyle{unsrt}
\bibliography{paper}

\end{document}